\documentclass[twocolumn,showpacs,preprintnumbers,floats,amsmath,amssymb]{revtex4}
\usepackage{graphicx}
\usepackage{dcolumn}
\usepackage{bm}
\begin{document}
\newcommand{\ds}{\displaystyle}
\newcommand{\be}{\begin{equation}}
\newcommand{\en}{\end{equation}}
\newcommand{\bea}{\begin{eqnarray}}
\newcommand{\ena}{\end{eqnarray}}
\topmargin -2cm
\title{Interacting cosmic fluids in power--law Friedmann-Robertson-Walker cosmological models}
\author{Mauricio Cataldo}
\altaffiliation{mcataldo@ubiobio.cl} \affiliation{Departamento de
F\'\i sica, Universidad del B\'\i o--B\'\i
o, Avenida Collao 1202, Casilla 5-C, Concepci\'on, Chile.\\}
\author{Patricio  Mella}
\altaffiliation{patriciomella@udec.cl} \affiliation{Departamento de F\'{\i}sica, Universidad de Concepci\'{o}n,\\
Casilla 160-C, Concepci\'{o}n, Chile. \\}
\author{Joel Saavedra}
\altaffiliation{Joel.Saavedra@ucv.cl} \affiliation{Instituto de F\'\i sica, Pontificia Universidad Cat\'olica de Valpara\'\i so, \\
Casilla 4950, Valpara\'\i so, Chile.}
\author{Paul  Minning}
\altaffiliation{pminning@udec.cl} \affiliation{Departamento de F\'{\i}sica, Universidad de Concepci\'{o}n,\\
Casilla 160-C, Concepci\'{o}n, Chile. \\}
\date{\today}
\begin{abstract}
{\bf {Abstract:}} We provide a detailed description for power--law
scaling Friedmann-Robertson-Walker cosmological scenarios
dominated by two interacting perfect fluid components during the
expansion. As a consequence of the mutual interaction between the
two fluids, neither component is conserved separately and the
energy densities are proportional to $1/t^{2}$. It is shown that
in flat FRW cosmological models there can exist interacting
superpositions of two perfect fluids (each of them having a
positive energy density) which accelerate the expansion of the
universe. In this family there also exist flat power law
cosmological scenarios where one of the fluids may have a
``cosmological constant" or ``vacuum energy" equation of state ($p
=-\rho$) interacting with the other component; this scenario
exactly mimics the behavior of the standard flat Friedmann
solution for a single fluid with a barotropic equation of state.
These possibilities of combining interacting perfect fluids do not
exist for the non-interacting mixtures of two perfect cosmic
fluids, where the general solution for the scale factor is not
described by power--law expressions and has a more complicated
behavior. In this study is considered also the associated single
fluid model interpretation for the interaction between two fluids.
\vspace{0.5cm} 
\end{abstract}
\smallskip\
\maketitle \section{Introduction}

Recent observational data give a strong motivation to study
general properties of Friedmann-Robertson-Walker (FRW)
cosmological models containing more than one fluid. The standard
modern cosmology considers the total energy density of the
Universe dominated today by the densities of two components: the
dark matter (which has an attractive gravitational effect like
usual matter), and the dark energy (a kind of vacuum energy with a
negative pressure)~\cite{Peebles}.


Usually the universe is modeled with perfect fluids and with
mixtures of non-interacting perfect fluids~\cite{Gunzig}. This
means that it is assumed that there is no conversion (energy
transfer) among the components and that each of them evolves
separately according to standard conservation laws. However, there
are no observational data confirming that this be the only
possible scenario. This means that we can consider plausible
cosmological models containing fluids which interact with each
other. In this case the transfers of energy among these fluids
play an important role. Thanks to these energy exchanges, in some
cosmological models it is possible, for example, to give a
reasonable explanation for the observed late acceleration of the
universe~\cite{Riess} and for the coincidence
problem~\cite{Guo,Berger}, since some mechanisms could exist for
converting one fluid into another. There are many other
cosmological situations where this exchange of energy was
considered. For example, the interaction between dust--like matter
and radiation was first considered by Tolman~\cite{Tolman} and
Davidson~\cite{Davidson}. It is interesting to note that Davidson
considered only positive pressures since at that time there was no
observational evidence for negative stresses in intergalactic
space. Also were considered cosmological models with decay of
massive particles into radiation, or with matter
creation~\cite{Lima}. For more examples see Barrow and
Clifton~\cite{Barrow}, and the cites contained therein.

On the other hand, for a long time cosmologists have used the most
simple  solutions of the Einstein field equations, applying them
to cosmology, and developing the so--called standard model. In
this sense, the main aim of this paper is to consider the most
simple non--trivial cosmological scenarios for an interacting
mixture of two cosmic fluids described by power--law scale
factors, i.e. the expansion (contraction) as a power--law in time.
In a general context the power--law cosmologies are defined by
their growth of the cosmological scale factor as $a(t) \propto
t^\alpha$. The observed expanding stage of the universe is
described by $\alpha > 0$; for $\alpha < 0$ we have a contracting
universe ($t>0$). The behavior of the universe in power--law
cosmologies is completely described by the Hubble parameter
$H=\dot{a}/a$ and the deceleration parameter
$q(t)=-\ddot{a}a/\dot{a}^2$. For $a(t)=t^\alpha$ it takes the form
$q_0=-(\alpha-1)/\alpha$ implying that the universe expands with a
constant velocity for $\alpha=1$ and with an accelerated expansion
for $\alpha > 1$ since, if the expansion is speeding up, the
deceleration parameter must be negative.

The interest in power law FRW cosmologies is not new. The
motivation for studying  this kind of cosmological scenarios comes
for example from the following aspects. There is good evidence for
such a power-law expansion during the radiation and matter
dominated epochs, for which $\alpha=1/2$ and $\alpha=2/3$
respectively, so in both cases we have $\alpha < 1$, implying that
these epochs had a decelerated expansion.

One may also consider a simple inflationary model characterized by
a period in which the scale factor is a power law in time with
$\alpha > 1$, which is called power law inflation~\cite{Lucchin}.
This occurs when the state parameter $\omega$ in the barotropic
equation of state $p = \omega \rho$ is constant and $\omega  < -
1/3$. Power-law inflationary models allow us to solve the horizon
and flatness problems, among others; however the main theoretical
problem which arises from these models is that inflation never
comes to an end because its slow-roll parameter is proportional to
$1/\alpha$ and then is constant~\cite{Marozzi}. Nevertheless its
advantage lies in the possibility of analytically computing the
solutions of the perturbation equations and the corresponding
power spectra~\cite{Rojas}.

On the other hand, it is interesting to note that there exist a
class of cosmological models, that attempt to dynamically solve
the Cosmological constant problem, in which the scale factor grows
as a power law in time, regardless of the matter content or
cosmological epochs~\cite{Weingberg}. Such power law cosmologies,
with $\alpha \thickapprox 1$, satisfy the observational
constraints on the present age of the Universe, the
magnitude--redshift relation of the type Ia supernova and the
angular size for a large sample of milliarcsecond compact radio
sources; however there are some inconsistencies with the
requirement that primordial nucleosynthesis produces light
elements in abundances consistent with those inferred from
observational data~\cite{Kaplinghat}.

Lastly, although there is no clear evidence for a pure power-law
expansion today, maybe the Universe has entered an epoch of
accelerated power--law expansion, or perhaps in the future it
could enter such an expansion, and this could imply that the
Universe will expand forever and never will exit from this stage.
In this case only cosmological scenarios with $\alpha > 1$ may
present a physical interest to us.

Another remarkable property of a power--law scale factor is that
in our study the mutual exchange of energy between two perfect
fluids can be described by energy densities which are proportional
to $1/t^2$ and the interaction term proportional to $1/t^3$. The
advantage of considering this kind of interacting fluids is that
the energy densities evolve at the same rate, so their ratio is a
constant quantity, thus satisfying the so--called cosmological
coincidence problem, namely: Why the matter and dark energy
densities are of the same order today?


\section{Field equations for two Interacting fluids}
For an open, closed or flat FRW universe filled with two fluids
$\rho_{_1 }$ and $\rho_{_{2}}$, the Friedmann equation is given by
\begin{eqnarray}\label{friedmann equation}
3 H^2+\frac{3k}{a^2}= \kappa\left(\rho_{_1 }+\rho_{_{2}}\right),
\end{eqnarray}
where $k=-1, 0, 1$ (from now on we shall set $\kappa=8 \pi G=1$).
We postulate that the two components $\rho_{_1 }$ and
$\rho_{_{2}}$ interact through the interaction term $Q$ according
to
\begin{eqnarray}\label{masQ}
\dot{\rho_{_1 }}+3 H \left(\rho_{_1 }+p_{_{1}}\right)=Q, \\
\dot{\rho}_{_{2}}+3 H \left(\rho_{_{2} }+p_{_{2}}\right)=-Q.\label{menosQ}
\end{eqnarray}
Note that if $Q>0$ we have that there exists a transfer of energy
from the fluid $\rho{_{_2}}$ to the fluid $\rho{_{_1}}$. The
nature of the $Q$ term is not clear at all. It may arise in
principle from some microscopic mechanisms~\cite{Berger}. For
solving these equations different forms for the interaction term
$Q$ have been considered.

If $Q=0$ we have two non--interacting fluids, and then each fluid
satisfies the standard conservation equation separately. Let us
consider the flat case, i.e. $k=0$. Putting $Q=0$ into
Eqs~(\ref{masQ}) and~(\ref{menosQ}) we have for each conserved
component that $\rho_1=C_1 a(t)^{-3(1+\omega_1)}, \rho_2=C_2
a(t)^{-3(1+\omega_2)}$. Since we are interested in power law
scenarios, the above energy densities take the following form:
$\rho_1=\rho_{10} t^{-3\alpha(1+\omega_1)}, \rho_2=\rho_{20}
t^{-3\alpha(1+\omega_2)}$, and from Friedmann
equation~(\ref{friedmann equation}) we obtain that
$\omega_1=\omega_2$. This implies that always both fluids have the
same equation of state and then the non--interacting superposition
of two fluids is really a trivial case in power law cosmologies.
However, as we shall see, the description of a superposition of
two interacting fluids is not at all trivial.

\subsection{Closed and open power--law interacting cosmologies}
Let us now consider FRW cosmologies with $k=-1,1$ filled with
interacting matter sources which satisfy a barotropic equation of
state, i.e
\begin{equation}\label{5}
p_{_1}=\omega_{_1} \rho{_{_1}}, p_{_2}=\omega_{_2} \rho{_{_2}},
\end{equation}
where $\omega_{_1}$, $\omega_{_2}$ are constant state parameters.
We shall define the scale factor as $a(t)=t^\alpha$, where
$\alpha$ is a constant parameter. This implies that $H=\alpha/t$
and, taking into account the curvature term $3k/a^2$ of
Eq.~(\ref{friedmann equation}), we conclude that $\alpha=1$, in
order to obtain energy density scales in the same manner as the
curvature term. Since  $a=t$ we have no acceleration and the
universe will either expand with constant velocity or collapse
with constant velocity.

This strictly linear evolution of the scale factor has been
considered before in the literature. For example in
Ref.~\cite{Dev} it is shown that an open FRW cosmology with a
linear evolution of the scale factor is consistent with the latest
SNe Ia observations and constraints arising from age of old
quasars.

Now, from Eq.~(\ref{friedmann equation}) and the resultant
equation from the addition of Eqs.~(\ref{masQ})
and~(\ref{menosQ}), we obtain that
\begin{eqnarray}\label{k1}
\rho_{k1}(t)=\frac{(1+k)(1+3\omega_{_2})}{(\omega_{_2}-\omega_{_1})\,
t^2}, \\ \label{k2}
\rho_{k2}(t)=-\frac{(1+k)(1+3\omega_{_1})}{(\omega_{_2}-\omega_{_1})\,
t^2},
\end{eqnarray}
and then the interacting term is given by
\begin{equation}\label{qk}
Q(t)=\frac{(3+k)(1+3\omega_{_1})(1+3\omega_{_2})}{3(\omega_{_2}-\omega_{_1})\,
t^3}.
\end{equation}
This term may be rewritten as
\begin{equation}\label{qk}
Q(t)=(1+3\omega_{_1}) \, H \, \rho_{k1}=-(1+3\omega_{_2}) \, H \,
\rho_{k2},
\end{equation}
which implies that the interacting term is proportional to the
expansion rate of the universe and to one of the individual
densities, so $Q \sim t^{-3}$.

\subsection{Flat Power--law interacting cosmologies}

Let us now consider interacting matter sources in flat FRW
universes satisfying the barotropic equations of state~(\ref{5}).
This means that we must put $k=0$ and $a(t)=t^\alpha$ into
Eq.~(\ref{friedmann equation})--(\ref{menosQ}). Taking into
account the Friedmann equation~(\ref{friedmann equation}) and the
resultant equation from the addition of Eqs.~(\ref{masQ})
and~(\ref{menosQ}) we conclude that the general solution is given
by
\begin{eqnarray}\label{rho10}
\rho_{_{1}}(t)=\frac{\rho_{_{10}}}{t^2}=    \frac{\alpha(-2+3
\alpha(1+\omega_{_2}))}{(\omega_{_2}-\omega_{_1}) \,t^2},
\end{eqnarray}
and
\begin{eqnarray}\label{rho20}
\rho_{_{2}}(t)=\frac{\rho_{_{20}}}{t^2}=\frac{\alpha(2-3
\alpha(1+\omega_{_1}))}{(\omega_{_2}-\omega_{_1}) \, t^2},
\end{eqnarray}
where the Q--term takes the form
\begin{eqnarray}
Q=\frac{\alpha(3 \alpha(1+\omega_{_1})-2)(3
\alpha(1+\omega_{_2})-2)}{(\omega_{_2}-\omega_{_1})\, t^3}.
\end{eqnarray}
This implies that the interaction term may be rewritten as
\begin{eqnarray}\label{QQ}
Q=\frac{(3 \alpha(1+\omega_{_1})-2)}{\alpha} \, H \, \rho_{_1} =
 -\frac{(3 \alpha(1+\omega_{_2})-2)}{\alpha} \, H \, \rho_{_2}.
 \nonumber \\
\end{eqnarray}
From this equation we conclude that, as before, the $Q$--term is
proportional to one of the individual densities and to the
expansion rate of the universe, so $Q \sim t^{-3}$.


\section{Specific two-fluid interactions}
Since we are primarily interested in a characterization of a
cosmological interaction between two fluids, we shall mainly
consider special assumptions in order to have some classification
schemes and detailed relationships between the power--law scale
factor and equations of state of the interacting cosmic fluids. In
this sense we shall consider that the weak energy condition (WEC)
holds and then we shall require the simultaneous fulfillment of
the conditions
$\rho_{_1} \geq 0, \rho_{_2} \geq 0$,
which implies that $\rho_{_{eff}}=\rho_{_1}+\rho_{_2} \geq 0$.
These conditions will imply some constraints on the state
parameters $\omega_{_1}$ and $\omega_{_2}$.

\subsection{Open and closed FRW universes}
In this section we first consider the case $k \neq 0$. It is clear
from Eqs.~(\ref{k1}) and~(\ref{k2}) that for $k=-1$ the energy
densities $\rho_{k1}$ and $\rho_{k2}$ vanish so, in this case of
an open FRW, it is not possible to have a cosmological evolution
with two interacting fluids. This kind of evolution is possible
only for a closed FRW Universe. Putting $k=1$ into Eqs.~(\ref{k1})
and~(\ref{k2}) and requiring the fulfillment of the conditions
$\rho_{_1} \geq 0, \rho_{_2} \geq 0$, we obtain the following
constraints on the state parameters:
\begin{eqnarray}\label{cqk1}
\omega_{_1} \leq -1/3, \omega_{_2}\geq -1/3,
\end{eqnarray}
for $\omega_{_2}>\omega_{_1}$ or, equivalently
$\omega_{_2} \leq -1/3, \omega_{_1}\geq -1/3$,
for $\omega_{_1}>\omega_{_2}$. From these expressions we conclude
that always one of the interacting fluids must be either a dark or
a phantom fluid. The constraints~(\ref{cqk1}) on the state
parameters imply that $Q<0$ (see Eq.~(\ref{qk})), so the energy is
transferred from a dark ($-1\leq\omega_{_1}\leq-1/3$) or a phantom
($\omega_{_1}<-1$) fluid to the other matter component whose state
parameter $\omega_{_2} > -1/3$.

Another aspect to be considered is the behavior of the constant
ratio of energies
$r_1=\rho_{12}/\rho_{11}=-\frac{1+3\omega_{_1}}{1+3\omega_{_2}}$
as a function of the model parameters $\omega_{_1}$ and
$\omega_{_2}$.
For cosmological scenarios which satisfy the requirement
$\omega_{_1}+\omega_{_2}<-2/3$, the matter component whose state
parameter $\omega_{_2}>-1/3$ dominates over the dark or phantom
fluid ($\omega_{_2}>\omega_{_1}$).

As here we have always a constant ratio for energy densities since
they are proportional to $1/t^2$, we shall use hereafter the word
``dominate" in the sense of ``the larger matter component of the
universe is". Note that this same ``dominant $=$ predominant"
component will always continue to be the largest one throughout
the cosmic evolution of the cosmological model.

\begin{figure}
\includegraphics{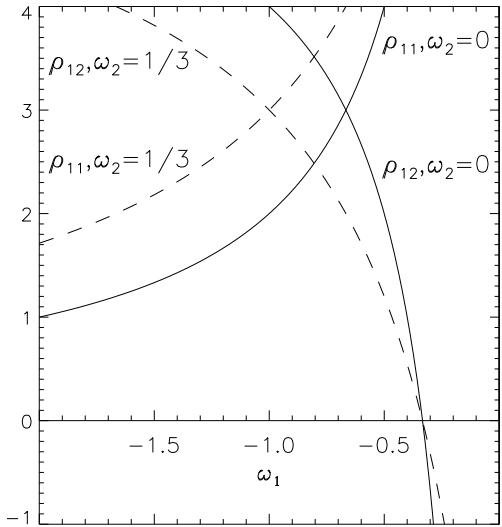}
\caption{\label{DensidadesClosedK} The behaviors of energy
densities $\rho_{k1}t^2$ and $\rho_{k2} \,t^2$ from
Eqs.~(\ref{k1}) and~(\ref{k2}) are plotted, for closed ($k=1$)
interacting FRW models, as functions of $\omega_{_1}$. In one case
(solid lines) we have taken $\omega_{_2}=0$ so that one of
interacting fluids is a dust distribution. There are dust
dominated cosmological scenarios for $\omega_{_1}< -2/3$. Note
that both energy densities are positive for $\omega_{_1}< -1/3$.
In another case (dashed lines) we have taken $\omega_{_2}=1/3$ so
that one of interacting fluids is a radiation distribution. There
are radiation dominated cosmological scenarios for $\omega_{_1}<
-1$. Note that both energy densities are positive also for
$\omega_{_1}< -1/3$.}
\end{figure}

As examples of the behavior of energy densities in some
interacting closed FRW cosmologies, Fig.~\ref{DensidadesClosedK}
is plotted. The interaction of dust with dark or with phantom
fluid is considered. We see that both energy densities are
positive for $\omega_{_1}< -1/3$ and the dark fluid dominates over
dust for $-2/3< \omega_{_1}< -1/3$. For $-1< \omega_{_1}< -2/3$
and for $\omega_{_1}< -1$ the dust distribution dominates over the
dark and the phantom fluids respectively. On the other hand, also
is considered the interaction of radiation with dark or with
phantom fluid. The energy densities are positive for $\omega_{_1}<
-1/3$. The dark fluid dominates over radiation for $-1<
\omega_{_1}< -1/3$ and, for $\omega_{_1}< -1$ the radiation
distribution dominates over the phantom fluid.

\subsection{Two---fluid interactions in flat FRW universes}
Now we shall study interacting fluids in flat FRW cosmologies,
i.e. $k=0$. In order to make the characterization of the
interaction between the two fluids we first consider the parameter
$\alpha$ to be a free one in Eqs.~(\ref{rho10}) and~(\ref{rho20}).
This means that we shall seek all possible cosmic expansion rates
for cosmological scenarios with fixed equations of state for the
two interacting fluids. Note that $\rho_{_1}=0$ if $\alpha=0$ and
if $\alpha=2/(3(1+\omega_{_2}))$; $\rho_{_2}=0$ if $\alpha=0$ and
if $\alpha=2/(3(1+\omega_{_1}))$.
So if we require simultaneously $\rho_{_{10}} \geq 0$ and
$\rho_{_{20}} \geq 0$, we  obtain the following possible
combinations. For $\omega_{_2}>\omega_{_1}$:
\begin{eqnarray}\label{COP1}
\frac{2}{3(1+\omega_{_2})}<\alpha < \frac{2}{3(1+\omega_{_1})},
(\omega_{_2}>\omega_{_1}>-1); \\ \label{COP1a} -\infty < \alpha <
\frac{2}{3(1+\omega_{_1})},
\frac{2}{3(1+\omega_{_2})}<\alpha<+\infty,
\nonumber \\
(\omega_{_1}<-1<\omega_{_2}); \\
\label{COP1b}\frac{2}{3(1+\omega_{_2})}< \alpha <
\frac{2}{3(1+\omega_{_1})}, (\omega_{_1}< \omega_{_2}<-1).
\end{eqnarray}
For $\omega_{_2}<\omega_{_1}$:
\begin{eqnarray}\label{COP2a}
\frac{2}{3(1+\omega_{_1})} < \alpha < \frac{2}{3(1+\omega_{_2})},
(-1<\omega_{_2}<\omega_{_1}); \\ \label{COP2b} -\infty<\alpha<
\frac{2}{3(1+\omega_{_2})}, \frac{2}{3(1+\omega_{_1})}<
\alpha<+\infty, \nonumber \\  (\omega_{_2}<-1<\omega_{_1}); \\
\frac{2}{3(1+\omega_{_1})}<\alpha<\frac{2}{3(1+\omega_{_2})},
(\omega_{_2}< \omega_{_1}<-1). \label{COP2}
\end{eqnarray}
Notice that Eqs.~(\ref{COP1}) and~(\ref{COP2a}) are valid for
configurations which include two interacting fluids obeying the
dominant energy condition (DEC), Eqs.~(\ref{COP1a})
and~(\ref{COP2b}) are valid for configurations where one
interacting fluid obeys DEC and the other is a phantom fluid, and
Eqs.~(\ref{COP1b}) and~(\ref{COP2}) are valid for the description
of two interacting phantom fluids.

Now we shall consider specific two--fluid interactions. It must be
noted that the relations~(\ref{COP1})--(\ref{COP2}) are valid for
$\omega_{_1} \neq -1$ (or $\omega_{_2} \neq -1$). At the end of
this section we will study configurations for which $\omega_{_1}=
-1$ (or $\omega_{_2}= -1$).

\subsubsection{Dust--Perfect fluid interaction ($\omega_{_1}=0$,
$\omega_{_2} \neq 0$)}

We shall begin with the consideration of the interaction of dust
with any other perfect fluid configuration. This means that we
must put $\omega_{_1}=0$ into Eqs.~(\ref{rho10})
and~(\ref{rho20}), while $\omega_{_2}$ is still a free parameter.
Thus we have for a dust (d) and a perfect fluid (pf) interacting
configurations
\begin{eqnarray}\label{rho10dust}
\rho_{d}=\frac{\alpha(-2+3 \alpha(1+\omega_{_2}))}{\omega_{_2} t^2},
\end{eqnarray}
and
\begin{eqnarray}\label{rho20dust}
\rho_{pf}=\frac{\alpha(2-3 \alpha)}{\omega_{_2} t^2},
\end{eqnarray}
with the equations of state $p_{d}=0$, $p_{pf}=\omega_{_2}
\rho_{pf}$. For the requirement of simultaneous fulfillment of the
conditions $\rho_{_{1}} \geq 0$ and $\rho_{_{2}} \geq 0$ we obtain
from Eqs. (\ref{COP1})--(\ref{COP2}) that the following
constraints must be satisfied:
\begin{eqnarray}
\frac{2}{3(\omega_{_2}+1)} < \alpha < \frac{2}{3}, (\omega_{_2}>0);
 \label{dust P positivo} \\
\frac{2}{3} < \alpha < \frac{2}{3(\omega_{_2}+1)}, (-1<\omega_{_2} <0);
 \label{dust P negativo}\\
\frac{2}{3} < \alpha < +\infty,
-\infty < \alpha < \frac{2}{3(\omega_{_2}+1)},  \nonumber \\
(-\infty < \omega_{_2} < -1) \label{dust phantom}.
\end{eqnarray}
Note that, for the above constraints, the specified values of
$\omega_{_2}$ imply that really $0< \alpha < 2/3$ (for
$\omega_{_2}>0$), $2/3< \alpha < \infty $ (for $-1<\omega_{_2}<
0$), and  $2/3 < \alpha < \infty$ or $-\infty < \alpha < 0$(for
$-\infty < \omega_{_2}<-1$).

\begin{figure}
\includegraphics{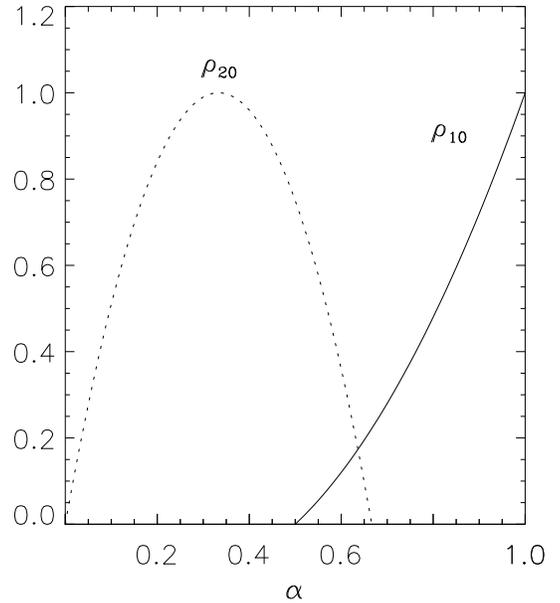}
\caption{\label{alpha graphic}  The behaviors of interacting dust
distribution  ($\rho_{_{1}} t^2=\rho_{_{10}}$) and radiation
distribution ($\rho_{_{2}} t^2=\rho_{_{20}}$) are plotted as
functions of $\alpha$. Both densities are positives in the
interval $\frac{1}{2}<\alpha<\frac{2}{3}$, and are equal at
$\alpha=\frac{7}{11}$. In the dust--radiation interacting case all
possible scenarios have a decelerated expansion and the whole
evolution is dominated by radiation if
$\frac{1}{2}<\alpha<\frac{7}{11}$, and is dominated by dust if
$\frac{7}{11}<\alpha<\frac{2}{3}$.}
\end{figure}

As a specific example we shall now consider in some detail the
dust--radiation interaction ($\omega_{_1}=0$, $\omega_{_2}=1/3$).
In this case we have
\begin{eqnarray}\label{dpf}
\rho_{_{1}}=\frac{\rho_{_{10}}}{t^2}=\frac{ 6\alpha (2
\alpha-1)}{t^2},  p_{_{1}}=0, \nonumber \\
\rho_{_{2}}=\frac{\rho_{_{20}}}{t^2}=\frac{3
\alpha(2-3\alpha)}{t^2}, p_{_{2}}=\frac{1}{3} \rho_{_{2}}.
\end{eqnarray}
In order to have simultaneously positive energy densities we must
require  that $1/2 <\alpha < 2/3$. The interaction term is given
by
$Q=\frac{3\alpha(3\alpha-2)(4\alpha-2)}{t^3}$.
For the interval $1/2 <\alpha < 2/3$ the Q--term is positive and
this means  that we have a transfer of energy from radiation to
the dust. In this scenario, for the value $1/2<\alpha=7/11 < 2/3$,
both densities are equal during all evolution. For the interval
$1/2<\alpha<7/11$ we have a radiation dominated universe, and for
$7/11<\alpha<2/3$ we have a dust dominated universe (see
Fig.\ref{alpha graphic}). In other words there exist
dust--radiation interacting cosmological scenarios dominated by
radiation or dust throughout all their evolution.

We want to remark that, in the case of non--interacting dust and
radiation, the expansion rate speeds up from $a(t)=t^{1/2}$ to the
$a(t)=t^{2/3}$ law. For the interacting dust--radiation case we
have a single expansion rate given by the $a(t)=t^\alpha$ law,
where $1/2 <\alpha < 2/3$. Thus in both cases the expansion of the
universe is decelerated.

\subsubsection{Phantom fluid--Perfect fluid interaction
($\omega_{_1}=-4/3$, $\omega_{_2} \neq 0$)}

On the other hand we shall now consider the interaction of a
phantom fluid with any other perfect  fluid configuration. We
choose as a representative cosmic fluid of phantom matter the
perfect fluid given by the equation of state $p=-4/3 \rho$. This
kind of perfect fluid was considered for example by the authors of
Ref.~\cite{Dabrowski}. This means that we must put
$\omega_{_1}=-4/3$ into the Eqs.~(\ref{rho10}) and~(\ref{rho20}),
while $\omega_{_2}$ is still a free parameter. Thus we have for a
phantom fluid (ph) and a perfect fluid (pf) interacting
configurations
\begin{eqnarray}\label{rho10ph}
\rho_{_{ph}}=\frac{3\alpha(-2+3
\alpha(1+\omega_{_2}))}{(3\omega_{_2}+4)t^2}, \\
\label{rho20ph}
\rho_{_{pf}}=\frac{3\alpha(2+\alpha)}{(3\omega_{_2}+4)t^2},
\end{eqnarray}
with the equations of state $p_{_{ph}}=-4/3 \, \rho_{_{ph}}$,
$p_{pf} =\omega_{_2} \rho_{pf}$. In order to have $\rho_{_{ph}}
\geq 0$ and $\rho_{_{pf}} \geq 0$ we obtain from Eqs.
(\ref{COP1})--(\ref{COP2}) that the following constraints must be
satisfied:
\begin{eqnarray}
-\infty < \alpha  < -2,  \frac{2}{3(\omega_{_2}+1)} < \alpha < \infty,
(\omega_{_2}>-1); \label{ph positivo} \\
\frac{2}{3(\omega_{_2}+1)} < \alpha < -2, (-4/3<\omega_{_2} <-1);
\label{ph negativo} \\
-2 < \alpha < \frac{2}{3(\omega_{_2}+2)}, (\omega_{_2} < -4/3).
\label{ph 1 negativo}
\end{eqnarray}
Note that, for the above constraints, the specified values of $
\omega_{_2}$  imply that really $-\infty< \alpha < -2$ or $0<
\alpha < \infty$ (for $\omega_{_2}>-1$), $-\infty < \alpha < -2$
(for $-4/3 < \omega_{_2}<-1$), and  $-2 < \alpha < 0$ for
($\omega_{_2}<-4/3$).

So for this cosmological scenario with a phantom fluid (given by
the  state parameter $\omega_{_1}=-4/3$) interacting with a
perfect fluid ($p_{pf}=\omega_{_2} \rho_{pf}$) the universe
expands only if $\omega_{_2}>-1$. In this case we can have
accelerated and non--accelerated expanding cosmologies.

\begin{figure}
\includegraphics{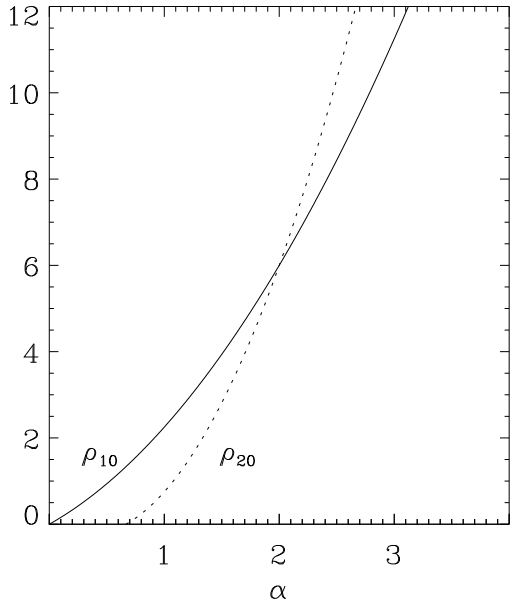}
\caption{\label{alpha graphic Phm} The behaviors of interacting
dust distribution ($\rho_{_{1}} t^2=\rho_{_{10}}$) and phantom
matter distribution ($\rho_{_{2} } t^2=\rho_{_{20}}$) are plotted
as functions of $\alpha$. In this case the interaction is
consistent with an expanding universe, and we have a
non--accelerated expansion for $2/3<\alpha<1$, and an accelerated
one for $1<\alpha<\infty$. For $\alpha=2$ we have
$\rho_{_{10}}=\rho_{_{20}}$, so if $2/3<\alpha<2$ the universe is
dominated by dust, and for $2<\alpha<\infty$ the universe is
dominated by the phantom matter component}
\end{figure}

As a specific example we shall consider the interaction of this
kind of phantom matter with a dust distribution. In this case we
have that $\omega_{_1}=-4/3$, $\omega_{_2}=0$ and then
$\rho_{_{1}}=\frac{\rho_{_{10}}}{t^2}=\frac{3 \alpha (3
\alpha-2)}{4 t^2}, p_{_{1}}=-\frac{4}{3} \rho_{_{1}},
\rho_{_{2}}=\frac{\rho_{_{20}}}{t^2}=\frac{3 \alpha(\alpha+2)}{4
t^2}, p_{_{2}}=0$.
In order to have simultaneously positive energy densities, we must
require that $\alpha < -2$ or $\alpha > 2/3$. The interaction term
is given by
$Q=\frac{3\alpha(\alpha+2)(2-3\alpha)}{4t^3}$.
For an interacting expansion, i.e. $\alpha > 2/3$, the Q--term is
positive and this means that we have a transfer of energy from
dust to the phantom matter. The interaction is consistent with an
expanding universe, and we have a non--accelerated expansion for
$2/3<\alpha<1$, and an accelerated one for $1<\alpha<\infty$. It
is interesting to note that, in this scenario, for $\alpha=2$ both
densities are equal and this implies that for $2/3<\alpha<2$ we
have scenarios where the universe is dominated by dust and, for
$2<\alpha<\infty$ we have cosmologies dominated by the phantom
matter component (see Fig.\ref{alpha graphic Phm}). In other words
there exist dust--phantom matter interacting cosmological
scenarios dominated by dust or by the phantom matter component
throughout all their evolution.

In conclusion, for the interacting dust--phantom matter case we
have  a single expansion rate given by the $a(t)=t^\alpha$ law,
where $\alpha < -2$ or $\alpha > 2/3$, and the expansion of the
universe may be decelerated or accelerated. Notice that this
result implies that we can have scenarios with $\alpha \gtrsim 1$
where the universe has an accelerated expansion but dust is
dominating over phantom matter.

\subsection{Interaction between effective ``vacuum
energy"and a perfect fluid}

As we stated above, the Eqs.~(\ref{COP1})--(\ref{COP2}) are valid
for $\omega_{_1} \neq -1$ and $\omega_{_2} \neq -1$. This means
that these equations can not be applied to interacting fluids with
an equation of state of the form $p_{_{1}}=-\rho_{_{1}}$ and
$p_{_{2}}=-\rho_{_{2}}$. However it is easy to see from
Eqs.~(\ref{rho10}) and~(\ref{rho20}) that the state parameters
$\omega_{_1}$ and $\omega_{_2}$ may take, although not
simultaneously, the value minus one.

Consider from now on in this section, that $\omega_{_1}= -1$.
Putting this value into Eqs.~(\ref{rho10}) and~(\ref{rho20}) we
obtain
\begin{eqnarray}\label{-1a}
\rho_{v}=\frac{\rho_{0v}}{t^2}=\frac{\alpha(-2+3
\alpha(1+\omega_{_2}))}{(1+\omega_{_2}) \, t^2},
\end{eqnarray}
and
\begin{eqnarray}\label{-1b}
\rho_{_{pf}}=\frac{\rho_{0pf}}{t^2}=\frac{2
\alpha}{(1+\omega_{_2}) \, t^2},
\end{eqnarray}
where the first fluid has a ``cosmological constant" or ``vacuum
energy" equation of state $p_{v}=-\rho_{v}$ and the second one is
a standard perfect fluid with an equation of state $p_{pf}
=\omega_{_2} \rho_{pf}$. The requirements that $\rho_{v} \geq 0$
and $\rho_{pf} \geq 0$ imply that
\begin{eqnarray}\label{1547}
\alpha > \frac{2}{3(1+\omega_2)}>0, (\omega_2 >-1); \\
\alpha < \frac{2}{3(1+\omega_2)}<0, (\omega_2 <-1).
\end{eqnarray}
It is interesting to note that the interaction of a perfect fluid
with a fluid with a ``cosmological constant" or ``vacuum energy"
equation of state exactly mimics the behavior of the standard
Friedmann solution for a single fluid with a barotropic equation
of state since for $-1<\omega_2<-1/3$ the expansion is accelerated
($\alpha
>1$), for $\omega_2 > -1/3$ we have decelerated expansion ($\alpha
< 1$), and for $\omega_2<-1$ we have that $-\infty < \alpha<0$.

In this case the interacting term is given by
\begin{eqnarray}
Q=\frac{2 \alpha(2-3 \alpha(1+\omega_{_2}))}{(1+\omega_{_2}) \,
t^3},
\end{eqnarray}
and we conclude that the interacting term is positive for
\begin{eqnarray}\label{1549}
0<\alpha<\frac{2}{3(1+\omega_2)}, (\omega_2>-1); \\
\frac{2}{3(1+\omega_2)}<\alpha<0, (\omega_2<-1).
\end{eqnarray}
From Eqs.~(\ref{1547}) and~(\ref{1549}) we obtain for fluids which
satisfy the DEC (i.e. $\omega_2>-1$) that the interacting term
$Q<0$, so that in the here considered interacting cosmological
scenarios always the energy is transferred from the effective
``vacuum energy" to the perfect fluid obeying the DEC.

\begin{figure}
\includegraphics{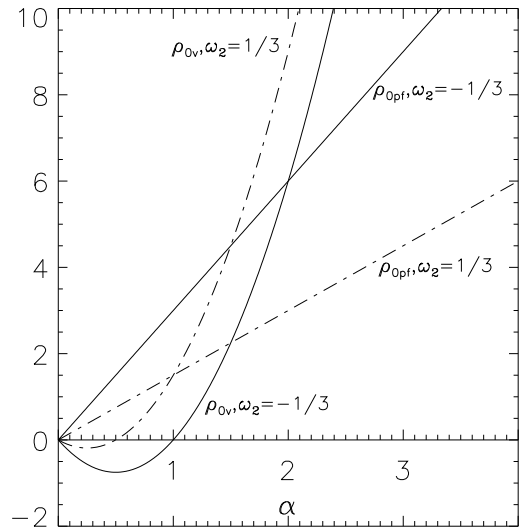}
\caption{\label{figratio1} In the figure is shown the behavior of
the energy densities of the interacting effective ``vacuum energy"
$\rho_{v}t^2$ and perfect fluid $\rho_{_{pf}}t^2$ for the extreme
cases $\omega_2=-1/3$ (solid lines) and $\omega_2=1/3$ (dashed
lines), see Eq.~(\ref{rr}). In the case of interaction between
effective ``vacuum energy" and radiation we see that the radiation
dominates only at stages with decelerated expansion ($1/2< \alpha
<1$). At $\alpha=1$ we have that $\rho_{v}=\rho_{_{pf}}$. For the
case $\omega_2=-1/3$ we have that the perfect fluid dominates over
the effective ``vacuum energy" at the range $1< \alpha <2$ so we
have accelerating expansion. For $\alpha >2$ the effective
``vacuum energy" dominates over the perfect fluid.}
\end{figure}

\begin{figure}
\includegraphics{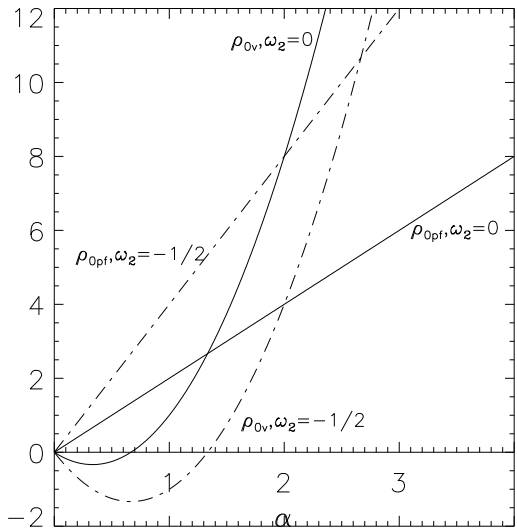}
\caption{\label{figratio15} In the figure is shown the behavior of
the energy densities of the interacting effective ``vacuum energy"
$\rho_{v}t^2$ and perfect fluid $\rho_{_{pf}}t^2$ for the cases
$\omega_2=-1/2$ (dashed lines) and $\omega_2=0$ (solid lines). In
the case of dust--effective ``vacuum energy" interaction we see
that dust dominates only in the range $2/3 < \alpha < 4/3$. It is
clear that for $1 <\alpha<4/3$ there is an accelerated expansion
dominated by dust. For the case $\omega_2=-1/2$ we have that the
perfect fluid dominates over the effective ``vacuum energy" at the
range $4/3< \alpha <8/3$ so we have accelerating expansion. For
$\alpha > 8/3$ the effective ``vacuum energy" dominates over the
perfect fluid.}
\end{figure}

Another aspect to be considered is the behavior of the constant
ratio of energies $r=\rho_{pf}/\rho_v=\frac{2}{3 \alpha
(1+\omega_2)-2}$ as function of the model parameters $\omega_2$
and $\alpha$.
It is easy to see that $r(\alpha,\omega_2)>1$ if
\begin{equation}\label{rr}
\frac{2}{3(1+\omega_2)} < \alpha < \frac{4}{3(1+\omega_2)}.
\end{equation}
So the perfect fluid dominates over the effective ``vacuum energy"
if, for a given $\omega_2$, the dimensionless parameter $\alpha$
varies in the specified above range. From Eq.~(\ref{rr}) we see
that, if $-1/3 <\omega_2 < 1/3$, there are cosmological scenarios
where the universe has accelerated and non--accelerated expansions
and is dominated by the perfect fluid. For $-1<\omega_2 < -1/3$
the Eq.~(\ref{rr}) implies that we have only accelerated scenarios
where the dark component dominates over the effective ``vacuum
energy"(see Fig.~\ref{figratio1} and Fig.~\ref{figratio15}). Note
that for $1/3<\omega_2 <1$ we can have decelerated expansion where
the effective ``vacuum energy" dominates over the perfect fluid
component.

\section{The effective fluid interpretation}
The main idea of this section is to study the conditions under
which  these two interacting sources are equivalent to an
effective fluid filling the universe. In other words we want to
associate an effective fluid interpretation with the interaction
of the two--fluid mixture. This can be made by associating with
the sum of pressures $p_{_1}$ and $p_{_2}$ an effective pressure
$p$, i.e.
\begin{eqnarray}\label{41}
p_{_1}+p_{_2}=\omega_{_1} \rho_{_1}+\omega_{_2} \rho_{_2}=p,
\end{eqnarray}
which has an equation of state given by
\begin{eqnarray}\label{42}
p=\gamma \rho=\gamma( \rho_{_1}+\rho_{_2}),
\end{eqnarray}
where $\gamma$ is a constant effective state parameter. Note that
the equation of state of the associated effective fluid is not
produced by physical particles and their
interaction~\cite{Gromov}.

In this sense for example, in the above discussed case of closed
FRW interacting cosmologies this single interpretation implies
that the effective fluid has an equation of state given by $p=-1/3
\,\rho$, and for the interaction between dust and radiation (see
Eqs.~(\ref{dpf})) this single fluid interpretation implies that
the effective fluid has an equation of state given by $p= \gamma
\rho$, where $0< \gamma < 1/3$. This means that the
dust--radiation interacting universe behaves as a FRW universe
filled with a single fluid with a state parameter varying in the
range $0< \gamma < 1/3$, preserving DEC.

Making some algebraic manipulations with Eqs.~(\ref{41})
and~(\ref{42}) we find that the effective state parameter $\gamma$
is related to the parameter $\alpha$ by
\begin{equation}\label{alpha gamma}
\gamma=\frac{2-3\alpha}{3 \alpha}.
\end{equation}
From this expression we see that the effective state parameter
$\gamma$  behaves as $\gamma \rightarrow -1$ for $\alpha
\rightarrow \pm \infty$. For $\alpha < 0$ we have the phantom
sector, since $\gamma < -1$.

Now we shall explore in more detail the effective interpretation
of the interacting two perfect fluids.
\subsection{Effective radiation and effective dust}
As we mentioned above we can associate a single fluid model with
the  interaction between two perfect fluids. In this section we
want to find all possible interacting superpositions for a given
$\alpha$--parameter. In order to do this we shall consider
$\alpha$ to be a given parameter, and $\omega_{_1}$ and
$\omega_{_2}$ to be free ones. This means that, for a fixed scale
factor (or Hubble parameter $H$), we shall find all possible state
equation configurations for each of the two interacting fluids.
From Eqs.~(\ref{rho10}) and~(\ref{rho20}) we shall obtain the
constraints on the free parameters $\omega_{_1}$ and
$\omega_{_2}$.

If we now require that $\rho_{_1} \geq 0$ and $\rho_{_2} \geq 0$
simultaneously, we obtain that
\begin{eqnarray}
\omega_{_1}<\frac{2-3 \alpha}{3 \alpha}< \omega_{_2}, (\alpha>0, \omega_{_2} > \omega_{_1}); \nonumber \\
\omega_{_1}<\frac{-2-3 |\alpha|}{3 |\alpha|}< \omega_{_2},
(\alpha<0, \omega_{_2} > \omega_{_1});
\end{eqnarray}
or equivalently $\omega_{_2}<\frac{2-3 \alpha}{3 \alpha}<
\omega_{_1} (\alpha>0)$ and $\omega_{_2}<\frac{-2-3 |\alpha|}{3
|\alpha|}< \omega_{_1} (\alpha<0$); for $\omega_{_2} <
\omega_{_1}$.

Here we have excluded the value $\omega_{_1}=2/3 \alpha-1$ (or
$\omega_{_2}=2/3 \alpha-1$) since this case gives a FRW universe
filled with a single fluid. From the above equations we see that,
for a physically plausible two-fluid interacting model associated
with a single effective model with equation of state $p=\gamma
\rho$ (see Eq.~(\ref{alpha gamma})) the whole ranges of validity
of the parameters $\omega_{_1}$ and $\omega_{_2}$ do not intersect
each other. If we want to have interacting perfect fluids which
obey the DEC, we constrain the parameters to the inequalities $-1
\leq \omega_{_1} \leq 1$ and $-1 \leq \omega_{_2} \leq 1$. In this
case one component (or both) may be a dark perfect fluid ($-1 \leq
\omega_{_i}\leq -1/3$, $i=1,2$). For the single effective model
the state parameter $\gamma$ also may obey the DEC $-1 \leq \gamma
\leq 1$. But we can consider more general situations. There are
interacting configurations where one fluid obeys DEC and the other
component does not (phantom fluid), but its interaction behaves
like a perfect fluid which obeys DEC. Note that this picture
completely excludes the possibility of having two interacting
phantom perfect fluids behaving like a fluid which obeys DEC.

As explicit examples we shall consider two interacting perfect
fluids which behave like either radiation, or dust or a kind of
phantom matter.

\subsubsection{Effective radiation fluid}  If the effective fluid
behaves like radiation ($\alpha=1/2, \, \gamma=1/3$), then the
free parameters ($\omega_{_2}>\omega_{_1}$) vary in the ranges
$-\infty < \omega_{_1}< 1/3$ and $1/3 < \omega_{_2}< \infty$. If
we require that the second fluid satisfies the DEC  (i.e. $1/3 <
\omega_{_2} \leq 1$), then we can consider its interaction with a
standard perfect fluid ($-1/3 < \omega_{_1}< 1/3$), or with a dark
fluid ($-1 \leq \omega_{_1}< -1/3$), or with a phantom fluid
($-\infty < \omega_{_1}< -1$). This model has a decelerated
expansion.

\subsubsection{Effective dust} If the effective fluid behaves like
dust ($\alpha=2/3, \, \gamma=0$), then the  free parameters
($\omega_{_2}>\omega_{_1}$) vary in the ranges $-\infty <
\omega_{_1}< 0$ and $0 < \omega_{_2}< \infty$. If we require that
the second fluid satisfies the DEC (i.e. $0 < \omega_{_2} \leq
1$), then we can consider its interaction with a standard perfect
fluid ($-1/3 < \omega_{_1}< 0$), or with a dark fluid ($-1 \leq
\omega_{_1}< -1/3$), or with a phantom fluid ($-\infty <
\omega_{_1}< -1$). This model has a decelerated expansion.

\subsubsection{An effective phantom fluid}
If the effective fluid behaves like a phantom one with state
parameter $\gamma=-4/3$ ($\alpha=-2$), then the free parameters
($\omega_{_2}>\omega_{_1}$) vary according to the ranges $-\infty
< \omega_{_1}< -4/3$ and $-4/3 < \omega_{_2}< \infty$. In this
case clearly we have the possibility of having an interacting
superposition of two phantom fluids. If we require that the second
fluid satisfies the DEC (i.e. $-1 < \omega_{_2} \leq 1$), then we
can consider its interaction with only a phantom fluid with state
parameter $-\infty < \omega_{_1}< -4/3$. In this case we always
have a contracting universe.

\section{Discussion}
In this paper we have provided a detailed description for
power--law  scaling cosmological models in the case of a FRW
universe dominated by two interacting perfect fluid components
during the expansion. We have shown that in this mathematical
description it is possible for each fluid component to require
that the conditions $\rho_{_1} \geq 0$ and $\rho_{_2} \geq 0$ may
be simultaneously fulfilled in order to have reasonable physical
values of state parameters $\omega_1$ and $\omega_2$ (we mean
either DEC, i.e. $-1\leq\omega_1,\omega_2\leq 1$; or else
$\omega_1,\omega_2<-1$). So from the required conditions we may
gain some insights for understanding essential features of
two--fluid interactions in power law FRW cosmologies. For example,
in the case of flat FRW universes, if we have a dust universe
(i.e. $a=t^{2/3}$) or a radiation universe ($a=t^{1/2}$), the
interacting fluids can not both be dark (or phantom) fluids. In
other words, ``dust" or ``radiation" effective universes can not
be filled with two interacting dark (or phantom) fluids.

On the other hand, we may apply our results to flat inflationary
cosmological models involving power law evolution for the scale
factor. This means that the parameter $\alpha$ must be constrained
to the range $\alpha > 1$, thus implying that any power law
inflationary model can be filled by two interacting fluids with
state parameters given by $\omega_1<-1/3< \omega_2$, so always one
of the interacting fluids must be either a dark fluid or a phantom
one.

One consequence of our results is that one may consider
accelerated cosmological models where one of the fluids is
described with the help of a minimally coupled scalar field which
interacts with a perfect fluid. Specifically, an exponential
potential may be used for the dark energy interacting component
which has a constant state parameter constrained to the range
$-1<\omega_1<-1/3$ provided that $\alpha
> 1$~\cite{Copeland}. In this case the scalar field evolves as
$\Phi \propto ln  \,  t$ and the perfect fluid has an equation of
state of the form $p=\omega_2 \rho$. Another possibility to be
considered is that the interacting dark energy component also may
be modelled as a rolling tachyon field. In general a rolling
tachyon condensate may be described by an effective fluid with
energy density and pressure given by
$\rho=V(T)/\sqrt{1-\dot{T}^2}$ and by $p=-V(T)\sqrt{1-\dot{T}^2}$,
where $T$ is the tachyon field and $V(T)$ is the tachyon
potential~\cite{Sami,Feinstein}. It is possible to obtain power
law inflationary cosmological models by assuming that the
potential is an inverse square in terms of the tachyon field, i.e.
$V(T)=\beta T^{-2}$, where $\beta>0$~\cite{Feinstein}. The same
fields may be considered for describing the present accelerating
stage of the universe. Notice that in this case it is also
possible to consider the interaction of a perfect fluid with
phantom energy ($\omega_1<-1$) in the form of an imaginary tachyon
field~\cite{Chimento}, which may be obtained by simply Wick
rotating the tachyon field~\cite{Setare}. A detailed analysis of
the here discussed ideas is in progress and will be published
elsewhere.

The here described variety of possibilities for combining
interacting perfect fluids with energy densities $\propto t^{-2}$
does not exist for the non-interacting mixtures of two perfect
cosmic fluids, where the general solution for the scale factor is
not described by power--law expressions and has a more complicated
behavior.

Note that the considered power--law cosmologies may describe
satisfactorily the interaction of dark matter (which is generally
assumed to be collisionless, i.e. described by a pressureless
fluid~\cite{Wang}) with any other perfect fluid configuration. So
the relations obtained in Section III for dust--perfect fluid
interaction may be applied to interacting dark matter.

It is interesting to observe that the here considered variety of
flat power--law scaling cosmological models is related to the
study made by Barrow and Clifton for cosmological models with a
mutual exchange of energy between two fluids at rates which are
proportional to a linear combination of their individual densities
and to the expansion rate of the universe~\cite{Barrow}. An
advantage of considering this type of interacting fluids is that
the energy densities at late times evolve at the same rate, so
their ratio is a constant quantity in agreement with the
so--called cosmological coincidence problem.

Specifically, for the kind of interaction studied by Barrow and
Clifton,  the power--law solutions behave at late times as
attractors of the general solution for the field
equations~(1)--(3) of Ref.~\cite{Barrow}. In particular, those
authors provided a simple mathematical description of the two
interacting fluids in an expanding flat FRW universe and showed
that the evolution can be reduced to a single nonlinear master
differential equation for the Hubble parameter $H$ of the form
$\ddot{H}+AH\dot{H}+BH^3=0$,
where $A$ and $B$ are
constants. This equation can be solved in physically relevant
cases and the authors provide an analysis of all possible
evolutions. Particular power--law solutions exist for the
expansion scale factor and are attractors at late times under
particular conditions. Note that the power--law scale factors are
solutions (self--similar solutions) for the master equation
$\ddot{H}+AH\dot{H}+BH^3=0$ with the parameters $A$ and $B$
constrained.

For the interacting flat cosmological scenarios discussed in our
paper, we see that Eq.~(\ref{QQ}) implies that the here considered
power--law cosmologies are the attractors for the particular
solution with $\alpha_{_{BC}}=\frac{(3
\alpha(1+\omega_{_1})-2}{\alpha}$, $\beta_{_{BC}}=0$ (or
$\alpha_{_{BC}}=0$, $\beta_{_{BC}}=\frac{(3
\alpha(1+\omega_{_2})-2)}{\alpha}$) of the above mentioned general
Barrow--Clifton solution, so all relations discussed in this work
may be applied to the late time behavior of this particular
solution and could help us clarify which kind of specific
interacting matter configurations are physically plausible today.

Lastly, today the observational data of Type Ia supernovae are
suggesting that our universe is undergoing accelerated
expansion~\cite{Riess}, so accelerated interacting superpositions
may play an important role in the study of two interacting fluids
at rates that are proportional to a linear combination of their
individual densities and to the expansion rate of the universe. On
the other hand, although there is no clear evidence for a pure
power-law expansion today, maybe the Universe has entered an epoch
of accelerated power--law expansion, or perhaps in the future it
could enter such an expansion, and this could imply that the
Universe will expand forever and never will exit from this stage.
From this point of view all found parameter constraints may shed
light on the possible cosmological scenarios to be considered. So
in this sense all interacting configurations with $0 < \alpha < 1$
could not represent interest today due to observational data.

\section{Acknowledgements}
This work was supported by Grants FONDECYT 1051086 (MC), 11060515
(JS), grant CONICYT 21070462 (PM), and by Direcci\'on de
Investigaci\'on de la Universidad del B\'\i o--B\'\i o (MC). The
financial support of Escuela de Graduados of the Universidad de
Concepci\'on is also acknowledged (PM).


\begin{thebibliography}{2}
\bibitem{Peebles} W.~M.~Wood-Vasey {\it et al.}  [ESSENCE Collaboration],
Astrophys.\ J.\  {\bf 666}, 694 (2007); P.~S.~Corasaniti, M.~Kunz,
D.~Parkinson, E.~J.~Copeland and B.~A.~Bassett, Phys.\ Rev.\  D
{\bf 70}, 083006 (2004); P.~J.~E.~Peebles and B.~Ratra, Rev.\
Mod.\ Phys.\  {\bf 75}, 559 (2003); S.~Perlmutter, M.~S.~Turner
and M.~J.~White, Phys.\ Rev.\ Lett.\  {\bf 83}, 670 (1999).
\bibitem{Gunzig} E. Gunzig, A.V. Nesteruk, M. Stokley, Gen. Rel. Grav.
{\bf 32}, 329 (2000);  M. Goliath, U.S. Nilsson, J. Math. Phys. {\bf 41}
6906 (2000); V.R. Gavrilov, V.N. Melnikov, S.T. Abdyrakhmanov Gen. Rel.
Grav.{\bf 36}, 1579 (2004); N. Pinto-Neto, E.S. Santini, F.T. Falciano
Phys. Lett. A {\bf 344}, 131 (2005); V. Bozza, G. Veneziano, JCAP {\bf
0509}, 007 (2005).
\bibitem{Riess} C. Armendariz-Picon, V.F. Mukhanov, P.J. Steinhardt,
Phys. Rev. Lett. {\bf 85}, 4438
(2000); R. Bean, J. Magueijo, Phys. Lett. B {\bf 517}, 177 (2001); A. G.
Riess et al., Astrophys. J. {\bf 607}, 665 (2004);
S.K. Srivastava, Phys. Lett. B {\bf 643}, 1(2006).
\bibitem{Guo} Z-K. Guo and Y-Z Zhang, Phys. Rev. D {\bf 71}, 023501 (2005);
D. Pavon and W. Zimdahl, Phys. Lett. B {\bf 628}, 206 (2005);
L.P. Chimento, A.S. Jakubi, D. Pavon, W. Zimdahl Phys. Rev. D {\bf 67}
083513, (2003).
\bibitem{Berger} M. S. Berger and H. Shojaei, Phys. Rev. D {\bf 73}, 083528
(2006).
\bibitem{Tolman} R. C. Tolman, Relativity, Thermodynamics and Cosmology
(Clarendon Press, London, 1934), section 165;
\bibitem{Davidson} W. Davidson, Mon. Not. R. Astron. Soc. {\bf 124}, 79 (1962);
\bibitem{Lima} J. A. S. Lima, Phys. Rev. D {\bf 54}, 2571 (1996); J. A. S.
Lima and M. Trodden, Phys. Rev. D {\bf 53}, 4280 (1996);
J. A. S. Lima, A. S. M. Germano, and L. R.W. Abramo, Phys. Rev. D
{\bf 53}, 4287 (1996).
\bibitem{Barrow} J. D. Barrow and T. Clifton, Phys. Rev. D {\bf 73},
103520 (2006).
\bibitem{Lucchin} L. F. Abbott, Nuc. Phys. B {\bf 244},
541 (1984); F. Lucchin and S. Matarrese, Phys. Rev. D {\bf 32},
1316 (1985).
\bibitem{Marozzi} G. Marozzi, Phys. Rev. D {\bf 76}, 043504 (2007).
\bibitem{Rojas} C. Rojas and V.M. Villalba, Phys. Rev. D
{\bf 75}, 063518 (2007); D.J. Liu and X.Z Li, Phys. Lett. B {\bf
600}, 1 (2004).
\bibitem{Weingberg} S. Weinberg, Rev. Mod. Phys.
{\bf 61} (1989); L.H. Ford, Phys Rev D {\bf 35}, 2339 (1987); A.D.
Dolgov, Phys. Rev. D {\bf 55}, 5881 (1997).
\bibitem{Kaplinghat} M. Kaplinghat, G. Steigman, I. Tkachev and
T.P. Walker, Phys. Rev. D {\bf 59}, 043514 (1999); M. Sethi, A.
Batra and D. Lohiya, Phys. Rev. D {\bf 60}, 108301 (1999); M.
Kaplinghat, G. Steigman and T.P. Walker, Phys. Rev. D {\bf 61},
103507 (2000); D. Jain, A. Dev and J.S. Alcaniz, Class. and
Quantum Grav. {\bf 20} 4485 (2003).
\bibitem{Dev} A. Dev, M. Safonova, D. Jain and D. Lohiya, Phys. \ Lett.\
B {\bf 548}, 12 (2002); G. Sethi, A. Dev and D. Jain, Phys.\
Lett.\  B {\bf 624}, 135 (2005).
\bibitem{Dabrowski} M.~Szydlowski and W.~Godlowski,
Phys.\ Lett.\  B {\bf 633}, 427 (2006); J. Kujat, R.J. Scherrer
and A.A. Sen, Phys.Rev. D {\bf 74},  083501 (2006); M. P.
Dabrowski, T. Stachowiak and M. Szydlowski, Phys. Rev. D {\bf 68},
103519 (2003).
\bibitem{Gromov} A. Gromov, Yu. Baryshev and P. Teerikorpi, Astron.
Astrophys. {\bf 415}, 813 (2004).
\bibitem{Copeland} E.~J.~Copeland, M.~Sami and S.~Tsujikawa, Int.\ J.\ Mod.\ Phys.\  D {\bf 15}, 1753
(2006).
\bibitem{Sami} M.~Sami, P.~Chingangbam and T.~Qureshi, Pramana {\bf 62}, 765 (2004).
\bibitem{Feinstein} A.~Feinstein, Phys.\ Rev.\  D {\bf
66}, 063511 (2002).
\bibitem{Chimento} L.~P.~Chimento and D.~Pavon, Phys.\ Rev.\  D {\bf 73}, 063511
(2006).
\bibitem{Setare} M.~R.~Setare, Phys.\ Lett.\  B {\bf 653}, 116
(2007); P.~F.~Gonzalez-Diaz, Phys.\ Rev.\  D {\bf 70}, 063530
(2004).
\bibitem{Wang} W. Wang, Y-X Gui, S-H. Zhang, G-H Guo and Y. Shao,
Mod.Phys. Lett. A {\bf 20}, 1443 (2005); C.M. Muller, Phys. Rev. D
{\bf 71}, 047302 (2005).
\end{thebibliography}
\end{document}